\documentclass[aps, prl,twocolumn]{revtex4}
\usepackage{color}
\usepackage{graphicx}
\usepackage{amsmath}
\usepackage{amssymb}
\usepackage{epstopdf}

\begin{document}
\title{Raman spectroscopy of graphene under ultrafast laser excitation}
\author{C. Ferrante$^{1,2,*}$, A. Virga$^{1,2}$, L. Benfatto$^3$, M. Martinati$^1$,\\  D. De Fazio$^4$, U. Sassi$^4$, C. Fasolato$^1$, A. K. Ott$^4$, P. Postorino$^{1}$,\\ D. Yoon$^4$, G. Cerullo$^5$, F. Mauri$^{1,2}$, A. C. Ferrari$^4$, T. Scopigno$^{1,2,*}$}
%\date{}
\affiliation{$^1$Dipartimento di Fisica, Universit\'a di Roma La Sapienza", I-00185, Roma, Italy}
\affiliation{$^2$Istituto Italiano di Tecnologia, Center for Life Nano Science @Sapienza, Rome, Italy}
\affiliation{$^3$Institute for Complex Systems, CNR, UoS Sapienza, I-00185, Rome, Italy}
\affiliation{$^4$Cambridge Graphene Centre,  University of Cambridge, Cambridge CB3 OFA,  UK}
\affiliation{$^5$IFN-CNR, Dipartimento di Fisica, Politecnico di Milano, P.zza L. da Vinci 32, 20133 Milano, Italy}
\affiliation{$^*$\textit{
 carino.ferrante@roma1.infn.it, tullio.scopigno@roma1.infn.it}}

\begin{abstract}
The equilibrium optical phonons of graphene are well characterized in terms of anharmonicity and electron-phonon interactions, however their non-equilibrium properties in the presence of hot charge carriers are still not fully explored. Here we study the Raman spectrum of graphene under ultrafast laser excitation with 3ps pulses, which trade off between impulsive stimulation and spectral resolution. We localize energy into hot carriers, generating non-equilibrium temperatures in the$\sim$1700-3100K range, far exceeding that of the phonon bath, while simultaneously detecting the Raman response. The linewidth of both G and 2D peaks show an increase as function of the electronic temperature. We explain this as a result of the Dirac cones' broadening and electron-phonon scattering in the highly excited transient regime, important for the emerging field of graphene-based photonics and optoelectronics.
\end{abstract}

\maketitle
The distribution of charge carriers has a pivotal role in determining fundamental features of condensed matter systems, such as mobility, electrical conductivity, spin-related effects, transport and optical properties. Understanding how these proprieties can be affected and, ultimately, manipulated by external perturbations is important for technological applications in diverse areas ranging from electronics to spintronics, optoelectronics and photonics\cite{Bonaccorso2010, Nanoscale2015, Koppens2014}.

The current picture of ultrafast light interaction with single layer graphene (SLG) can be summarized as follows\cite{Brida2013}. Absorbed photons create optically excited electron-hole (e-h) pairs. The subsequent relaxation towards thermal equilibrium occurs in three steps. Ultrafast electron-electron (e-e) scattering generates a hot Fermi-Dirac distribution within the first tens fs\cite{Tomadin2013}. The distribution then relaxes due to scattering with optical phonons (electron-phonon coupling), equilibrating within a few hundred fs\cite{Lazzeri2005, Butscher2007}. Finally, anharmonic decay into acoustic modes establishes thermodynamic equilibrium on the ps timescale\cite{Lui2010, Wu2012, Bonini2007}.

Raman spectroscopy is one of the most used characterization techniques in carbon science and technology\cite{FerrRob2004}. The measurement of the Raman spectrum of graphene\cite{FerrariPRL2006} triggered a huge effort to understand phonons (ph), e-ph, magneto-ph, and e-e interactions in graphene, as well as the influence of the number and orientation of layers, electric or magnetic fields, strain, doping, disorder, quality and types of edges, and functional groups\cite{FerrariNN2013, Malard2009, Froehlicher2015}. The Raman spectra of SLG and few layer graphene (FLG) consist of two fundamentally different sets of peaks. Those, such as D, G, 2D, present also in SLG, and due to in-plane vibrations\cite{FerrariPRL2006}, and others, such as the shear (C) modes\cite{Tan2012} and the layer breathing modes\cite{Sato2011,Lui2012} due to relative motions of the planes themselves, either perpendicular or parallel to their normal. The G peak corresponds to the high frequency \textit{E}$_{\mathrm{2g}}$ phonon at $\Gamma$. The D peak is associated to the ring breathing mode, and requires the presence of a defect for momentum conservation\cite{Tuinstra1970, Thomsen2000, FerrariPRB2000}. The 2D peak is the D overtone, it is always allowed as momentum conservation is satisfied in this case by two phonons with opposite wavevectors \cite{FerrariPRL2006}. Both D and 2D are activated by a double resonance (DR) mechanism, and are dispersive in nature due to a Kohn Anomaly at \textbf{K}\cite{Piscanec2004}.

Raman spectroscopy is usually performed under continuous wave (CW) excitation, therefore probing samples in thermodynamic equilibrium. The fast e-e and e-ph non-radiative recombination channels establish equilibrium conditions between charge carriers and lattice, preventing the study of the vibrational response in presence of an hot e-h population. Using an average power comparable to CW illumination (a few mW), ultrafast optical excitation can provide large fluences ($\sim1-15$J$/$m$^2$ at MHz repetition rates) over sufficiently short timescales (0.1-10ps) to impulsively generate a strongly out-of-equilibrium distributions of hot e-h pairs\cite{Lui2010, Yan2009, Breusing2011,Brida2013}. The potential implications of coupled electron and phonon dynamics for optoelectronics were discussed for nanoelectronic devices based on CW excitation\cite{Chae2010,PhysRevLett.104.227401, PhysRevB.93.075410, Mogulkoc201485, kim_bright_2015}. However, understanding the impact of transient photoinduced carrier temperatures on the colder SLG phonon bath is important for mastering out of equilibrium e-ph scattering, critical for photonics applications driven by carrier relaxation, such as ultrafast lasers\cite{sun2010}, detectors\cite{Bonaccorso2010, Koppens2014} and modulators\cite{Liu2011}. E.g, SLG can be used as a much broader-band alternative to semiconductors saturable absorbers\cite{sun2010}, for mode-locking and Q switching\cite{sun2010,Bonaccorso2010}.

\begin{figure*}
	\centerline{\includegraphics[width=160mm]{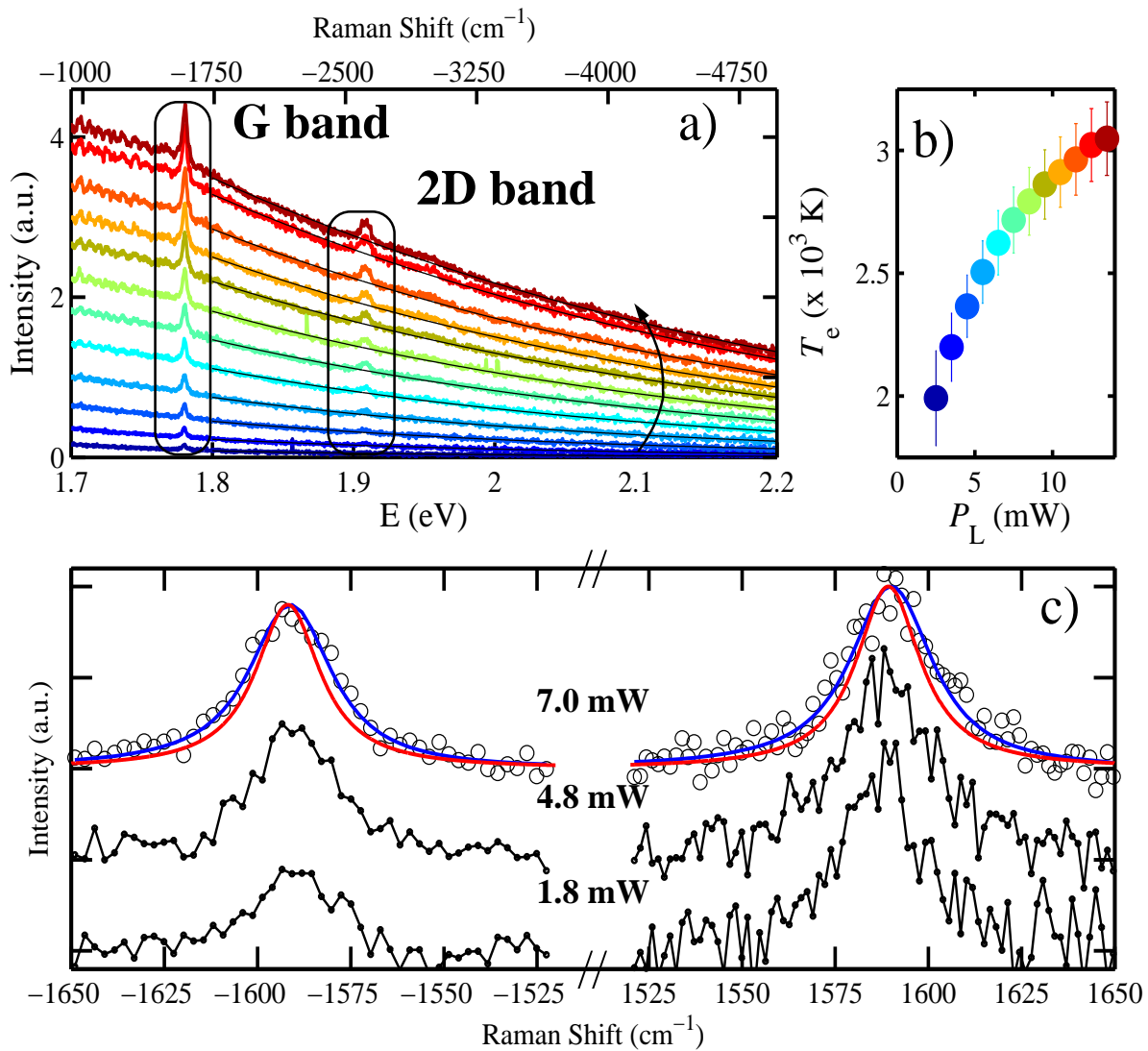}}
	\caption{\textbf{Spectral response of SLG.} a) AS Raman spectra under ultrafast excitation for laser powers increasing along the arrow direction. The $P_L$-dependent background is fitted by thermal emission (Eq.\ref{eqS1}, black lines) resulting in $T_e$ in the 1700-3100K range. b) T$_e$ as a function of P$_L$. c) Background subtracted, AS and S G peak (in black, normalized to the corresponding Stokes maximum) measured as function of $P_L$ in the range $\sim 1.8 \div 7.0$mW (corresponding to $T_e \sim 2000 \div 2700$K). Three representative $P_L$ values are shown. Best fits of the G peak (blue line), obtained as a convolution of a Lorentzian (red line) with the IRF are also reported for the largest $P_L$ value.}
	\label{fig1}
\end{figure*}

Here we characterize the optical phonons of SLG at high electronic temperatures $T_e$ by performing Raman spectroscopy under pulsed excitation. We use a 3ps pulse to achieve a trade off between the
narrow excitation bandwidth required for spectral resolution ($\frac{\delta \nu}{c}\leq$10cm$^{-1}$, being $\nu$[Hz] the laser frequency and c the speed of light, a condition met under CW excitation) and a pulse duration, $\delta t$, sufficiently short ($\delta t\leq$10ps, achieved using ultrafast laser sources) to generate an highly excited carrier distribution over the equilibrium phonon population, being those two quantities Fourier conjugates\cite{papoulis_1962} ($\frac{\delta\nu \cdot\delta t}{c} \geq 14.7$cm$^{-1}ps$). This allows us to determine the dependence of both phonon frequency and dephasing time on the hot carriers temperature, which we explain by a broadening of the Dirac cones.

\begin{figure*}
	\centerline{\includegraphics[width=150mm]{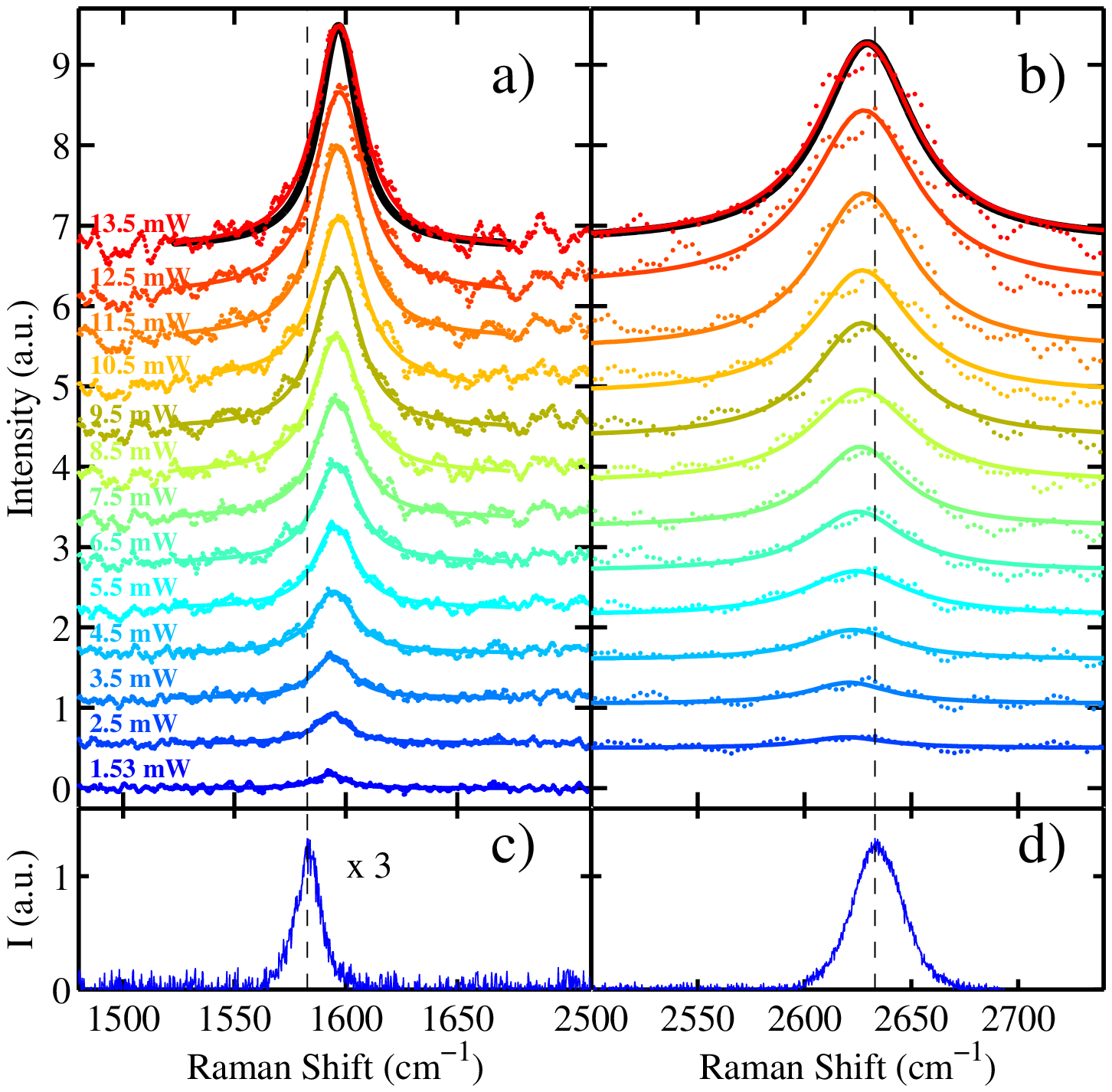}}
	\caption{\textbf{Raman spectra at different laser power.} (a) AS G and (b) 2D peak as function of $P_L$. (dots) Experimental data. (Lines) fitted Lorentzians convoluted with the spectral profile of the excitation pulse. The vertical dashed lines are the equilibrium, RT, Pos(G) and Pos(2D). (c) RT CW S G and (d) 2D peaks. The CW 2D is shifted by 5.4cm$^{-1}$ for comparison with the AS ps-Raman, see Methods. The relative calibration accuracy is $\sim$ 2cm$^{-1}$.}
	\label{fig2}
\end{figure*}

\section*{Results}
Fig.\ref{fig1}a plots a sequence of AntiStokes (AS) Raman spectra of SLG following ultrafast excitation at 1.58eV, as a function of excitation power $P_L$. The broad background stems from hot photoluminescence (PL) due to the inhibition of a full non-radiative recombination under high excitation densities\cite{freitag_2010,Lui2010, Stohr2010, PhysRevLett.104.227401}. This process, absent under CW excitation in pristine SLG\cite{Gokus2009}, is due to ultrafast photogeneration of charge carriers in the conduction band, congesting the e-ph decay pathway, which becomes progressively less efficient with increasing fluence. This non equilibrium PL recalls the grey body emission and can be in first approximation described by Planck's law\cite{freitag_2010, Lui2010, PhysRevLett.104.227401, kim_bright_2015}:
\begin{equation}
I(\hbar\omega, T_{e})=\mathcal{R}(\hbar\omega)\tau_{em}\eta \frac{\hbar\omega^3}{2\pi^2c^2}\left(e^{\frac{\hbar \omega}{kT_{e}}}-1\right)^{-1}
\label{eqS1}
\end{equation}
where $\eta$ is the emissivity, defined as the dimensionless ratio of the thermal radiation of the material to the radiation from an ideal black surface at the same temperature as given by the Stefan-Boltzmann law\cite{callen_1985}, $\tau_{em}$ is the emission time and $\mathcal{R}(\hbar\omega)$ is the frequency-dependent, dimensionless responsivity of our detection chain\cite{princeton, princeton2}. Refs.\cite{kim_bright_2015,freitag_2010,Lui2010} reported that, although Eq.\ref{eqS1} does not perfectly reproduce the entire grey body emission, the good agreement on a$\sim0.5$eV energy window is sufficient to extract $T_e$. By fitting the backgrounds of the Raman spectra with Eq.\ref{eqS1} (solid lines in Fig.\ref{fig1}a) we obtain $T_e$ as a function of $P_L$. Fig.\ref{fig1}b shows that $T_e$ can reach up to 3100K under our pulsed excitation conditions.

An upper estimate for the lattice temperature, $T_l$, can be derived assuming a full thermalization of the optical energy between vibrational and electronic degrees of freedom, i.e. evaluating the local equilibrium temperature, $T_{eq}$, by a specific heat argument (see Methods). We get $T_{eq}(P_{max})\sim680$K at the maximum excitation power, $P_{max}=13.5$mW. This is well below the corresponding $T_e$, indicating an out-of-equilibrium distribution of charge carriers. Thus, over our 3ps observation timescale, $T_l$ is well below $T_{eq}$.

Fig.\ref{fig1}c plots the AS and S G peaks, together with fits by Lorentzians (blue lines) convoluted with the laser bandwidth ($\sim9.5$cm$^{-1}$) and spectrometer resolution ($\sim6$cm$^{-1}$), which determine the instrumental response function, IRF (see Methods). The S data have a larger noise due to a more critical background subtraction, which also requires a wider accessible spectral range (see Methods). For this reason, we will focus on the AS spectral region, with an higher spectrometer resolution (1.2 cm$^{-1}$), Fig.\ref{fig2}. We obtain a full width at half maximum of the G peak, FWHM(G)$\sim21$cm$^{-1}$, larger than the CW one ($\sim 12.7$cm$^{-1}$). Similarly, we get FWHM(2D)$\sim$50-60cm$^{-1}$ over our $P_L$ range, instead of FWHM(2D)$\sim29$cm$^{-1}$ as measured on the same sample under CW excitation. To understand the origin of such large FWHM(G) and FWHM(2D) in pulsed excitation, we first consider the excitation power dependence of the SLG Raman response in the $1.53-13.5$mW range (the lower bound is defined by the detection capability of our setup). This shows that the position of the G peak, Pos(G), is significantly blueshifted (as reported for graphite in Ref.\cite{Yan2009}), while the position of the 2D peak, Pos(2D), is close to that measured under CW excitation, while both FWHM(G) and FWHM(2D) increase with $P_L$. Performing the same experiment on Si proves that the observed peaks broadening is not limited by our IRF (see Methods). Moreover, even the low resolution S data of the G band, collected in the range 1.8-7.0mW (a selection is shown in Fig.\ref{fig1}c), display a broadening ($(8\pm 4)10^{-3}$  cm$^{-1}$/K) and upshift ($(2.8\pm1.4)10^{-3}$  cm$^{-1}$/K), which is compatible with that of the high resolution AS measurements (Fig. \ref{fig3}d-e), $(7.4\pm0.5)10^{-3}$ cm$^{-1}$/K and $(3.2\pm0.2)10^{-3}$ cm$^{-1}$/K, respectively.

We note that phonons temperature estimates based on the AS/S intensity ratio\cite{Schomacker1986,FilhoPRB2001} (corrected for the wavelength dependent grating reflectivity and CCD efficiency) are hampered in graphene by two concurring effects. First, SLG's resonant response to any optical wavelength gives a non trivial wavelength dependent Raman excitation profile which modifies the Raman intensities with respect to the non-resonant case. Consequently, the AS/S ratio is no longer straightforwardly related to the thermal occupation \cite{goldstein2016raman}.
Second, in graphene one S created phonon may be subsequently annihilated by a correlated AS event. Although a complete theoretical description for this phenomenon is still laking, in practice, it results into an extra pumping in the AS side which does not allow to relate in the standard way AS/S ratio and phonon temperature via the thermal occupation factor \cite{ParraMurilloPRB2016}. Accordingly, the AS/S ratio approaching one at the largest excitation power in Fig.\ref{fig1}c (black circles) does not necessary imply a large increase of the G phonon temperature.

\section*{Discussion}

\begin{figure}
	\centerline{\includegraphics[width=90mm]{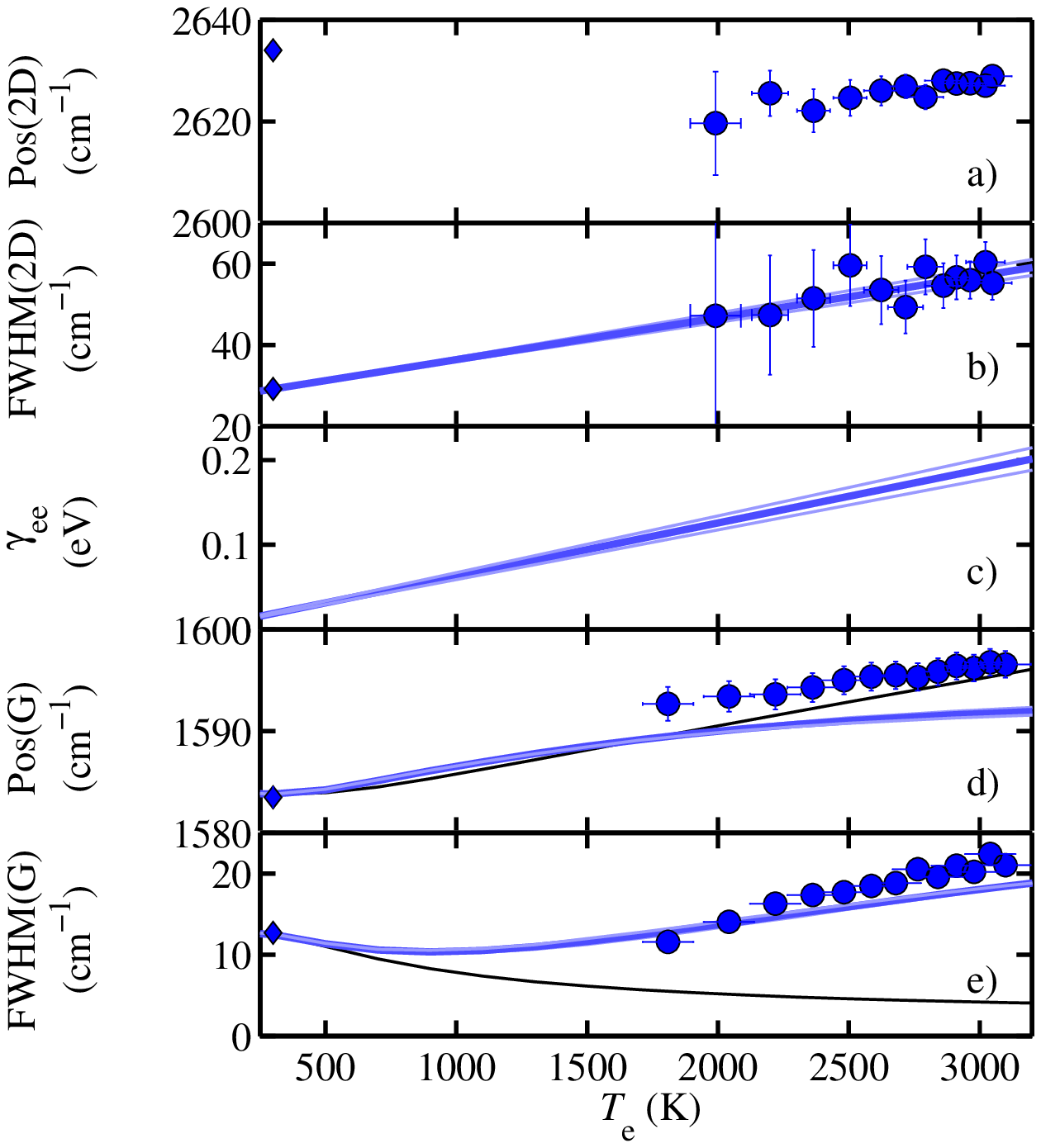}}
	\caption{\textbf{Comparison between theory and experiments.} a) Pos(2D), b) FWHM(2D), d) Pos(G), e) FWHM(G) as a function of $T_e$ for ps-excited Raman spectra. Solid diamonds in a,b,d,e represent the corresponding CW measurements. FWHM(2D) are used to determine the e-e contribution ($\gamma_{ee}$) to the Dirac cones broadening, shown in (c) (blue lines). Pos(G) and FWHM(G) are compared with theoretical predictions accounting for e-ph interaction in presence of electronic broadening (an additional RT anharmonic damping$\sim$2cm$^{-1}$\cite{Bonini2007} is included in the calculated FWHM(G)). Black lines are the theoretical predictions for $\gamma_{ee}=0$eV, while blue lines take into account an electronic band broadening linearly proportional to $T_e$ ($\gamma_{ee}=\alpha_e k_B T_e$). From the fit of $\gamma_{ee}$ in (c), we get $\frac{\alpha_e k_B}{hc}=0.51$cm$^{-1}$/K (thickest blue line). Values of $\frac{\alpha_e k_B}{hc}=0.46, 0.55$ cm$^{-1}$/K, corresponding to 99\% confidence boundaries, are also shown (thin light blue lines).}
	\label{fig3}
\end{figure}

\begin{figure*}
	\centerline{\includegraphics[width=170mm]{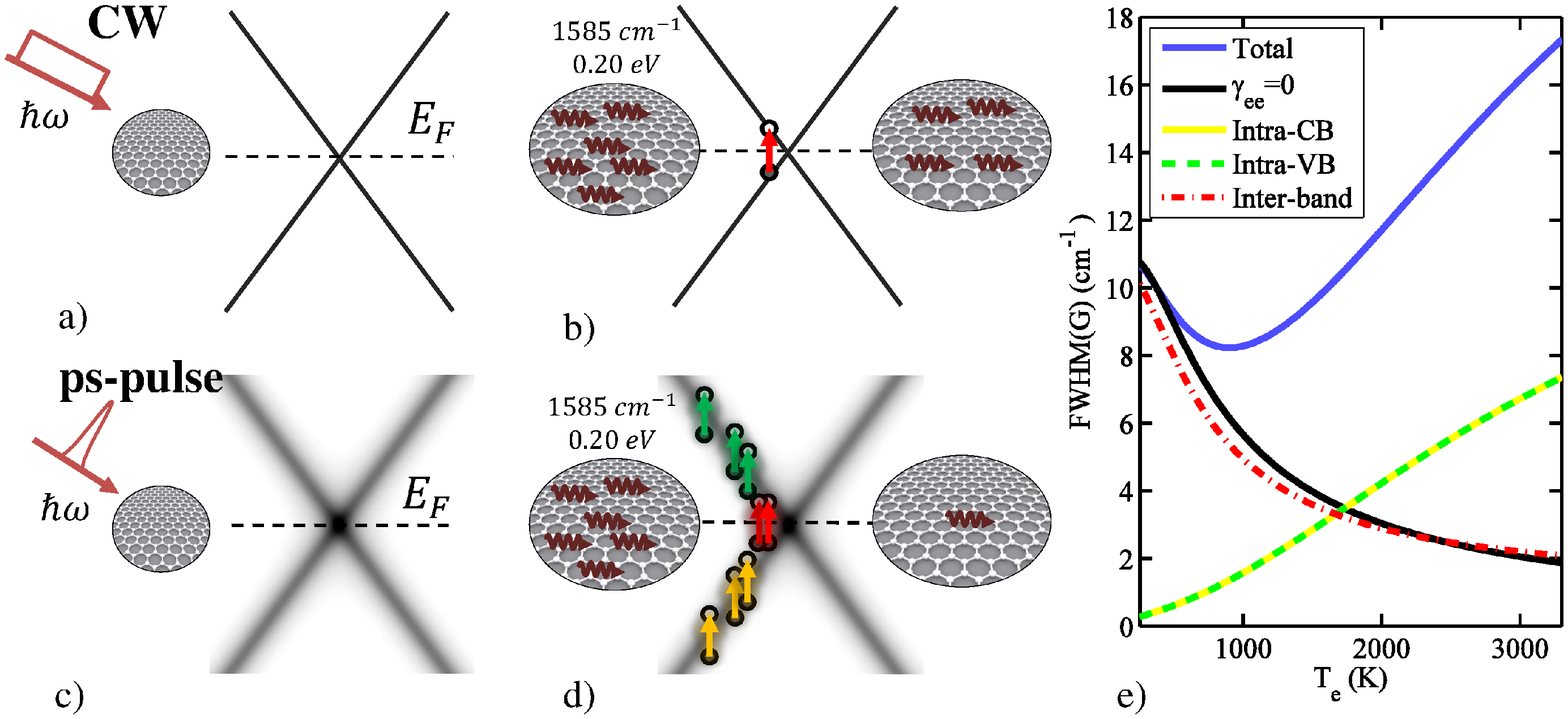}}
	\caption{\textbf{Effect of Dirac cone broadening on Raman process.} (a) CW photo-excitation with mW power does not affect the Dirac cone. (b) Accordingly, e-h formation induced by e-ph scattering only occurs in presence of resonant phonon excitation. (c) Under ps excitation, with average $P_L$ comparable to (a), the linear dispersion is smeared by the large $k_B T_e \approx \hbar \omega_G = 0.2$eV.  (d) Consequently, e-h formation is enhanced by the increased phonon absorption cross section, due to new intraband processes. (e) Corresponding contributions to FWHM(G) for the broadened inter-bands and intra-band processes for $\alpha_e k_B=0.51$ cm$^{-1}$/K.}
	\label{fig4}
\end{figure*}

Fig.\ref{fig3} plots Pos(2D), FWHM(2D), Pos(G), FWHM(G) as a function of $T_e$, estimated from the hot-PL. A comparison with CW measurements (633nm) at room temperature (RT) is also shown (blue diamonds). Under thermodynamic equilibrium, the temperature dependence of the Raman spectrum of SLG is dominated by anharmonicity, which is responsible for mode softening, leading to a redshift of the Raman peaks\cite{Apostolov,Basko2008, Bonini2007}. This differs from our experiments (Figs.4a-d), in which Pos(G) has an opposite trend (blue shift), and Pos(2D) is nearly $T_e$-independent, in agreement with Density Functional Perturbation Theory (DFPT) calculations, giving $\Delta$Pos(2D)$\sim$5cm$^{-1}$ in the range $T_e=300-3000$K (see Methods). This indicates the lack of significant anharmonic effects and suggests a dominant role of e-ph interaction on FWHM(G) and Pos(G), in the presence of a cold phonon bath at constant $T_l$ decoupled from the (large) $T_e$.

To derive the temperature dependence of such parameters, we first compute the phonon self-energy $\Pi(q=0,\omega_G^0)$, as for Refs.\cite{Piscanec2004,LazzeriPRL06, Pisana2007}:
\begin{multline}
\Pi(q=0,\omega^0_G,T_e)=\xi  \int_{0}^{\tilde\epsilon} d\epsilon \, \epsilon  \int_{-\infty}^{+\infty} dz \, dz'\sum_{s,s'} \\M_s(z,\epsilon)M_{s'}(z',\epsilon) \bigg[\frac{f_F(z-E_F)-f_F(z'-E_F)}{z-z'-\hbar\omega_G^0+\mathrm{i} \delta}\bigg]
\label{EqSE}
\end{multline}
Here $\xi= g^2/(2\hbar m_a \omega_G^0 v_F^2)=4.43 \times 10^{-3}$ is a dimensionless constant, $v_F$ is the Fermi velocity, $\tilde\epsilon$ is the upper cutoff of the linear dispersion $\epsilon=v_F k$, $m_a$ is the  carbon atom mass, $\hbar\omega_G^0=0.196$eV the bare phonon energy, $\delta$ is a positive arbitrary small number ($<4$meV), $g\sim12.3$eV is proportional to the e-ph coupling (EPC) \cite{Piscanec2004, Lazzeri2005, LazzeriPRL06,Ferrari200747}, $z$, $z'$ are the energy integration variables and  $f_F(z-E_F)$ is the Fermi-Dirac distribution with $E_F$ the Fermi energy. Although our samples are doped, $E_F$ significantly decreases as a function of $T_e$\cite{Chae2010}. Hence, we assume $E_F=0$ in the following calculations. The two indexes $s,s' = \mp 1$ denote the e and h branches, and $M_s(z,\epsilon)$ is the corresponding spectral function, which describes the electronic dispersion.

The self-energy expressed by Eq.\ref{EqSE} renormalizes the phonon Green's function according to the Dyson's equation\cite{Ando2006}:
\begin{equation}
D(\omega)=\frac{2\hbar\omega_G^0}{(\hbar\omega+i\delta)^2-(\hbar\omega_G^0)^2-2\hbar\omega_G^0\Pi(\omega)}
\end{equation}
so that the shift $\Delta$Pos(G) and FWHM(G) can be written as:
\begin{equation}
\begin{split}
\Delta \textnormal{POS(G)}&=\frac{1}{h c}\mathrm{Re} \left[\Pi(0,\omega_G^0,T_e)-\Pi(0,\omega_G^0,T_e=0)\right]\\
\textnormal{FWHM(G)}&=-\frac{2}{h c}\mathrm{Im}\Pi(0,\omega_G^0,T_e)
\end{split}
\end{equation}
where $h$ is the Planck constant.
FWHM(G) can be further simplified since the evaluation of $\mathrm{Im}\Pi(0,\omega_G^0,T_e)$ leads to $\delta(z-z'-\hbar\omega_G^0)$ in Eq.\ref{EqSE}, so that we get:
\begin{multline}
\textnormal{FWHM(G)}=\frac{\pi\xi}{h c}  \int_{0}^{\tilde\epsilon} d\epsilon \, \epsilon  \int_{-\infty}^{+\infty} dz\sum_{s,s'}\\ M_s(z,\epsilon)M_{s'}(z-\hbar\omega_G^0,\epsilon) \bigg[f_F(z)-f_F(z-\hbar\omega_G^0)\bigg]
\label{eqg}
\end{multline}
In the limit of vanishing broadening of the quasiparticle state, the SLG gapless linear dispersion is represented by the following spectral function\cite{Ando2006}:
\begin{equation}
\label{ms}
M_s(z,\epsilon)=\delta(z+s\epsilon), \quad s=\pm1,
\end{equation}
This rules the energy conservation in Eq.\ref{eqg} and allows only transitions between h and e states with energy difference $2\epsilon=\hbar\omega_G^0$. Thus, we get\cite{Piscanec2004, LazzeriPRL06, Pisana2007}:
\begin{multline}
\label{eqclean}
\textnormal{FWHM(G)}=\textnormal{FWHM(G)}^0 \left[f_F(-\hbar\omega_G^0/2) -f_F(\hbar\omega_G^0/2)\right]
\end{multline}
where $\textnormal{FWHM(G)}^0=\frac{\pi \xi \hbar \omega_G^0}{2 h c}\sim11$cm$^{-1}$\cite{Bonini2007}. This value, with the additional$\sim$2cm$^{-1}$ contribution arising from anharmonic effects\cite{Bonini2007}, is in agreement with the CW measurement at $T_e=T_{eq}=300$K (see diamond in Fig.\ref{fig3}e) corresponding to fluences$\ll 1$J/m$^2$. Eq.\ref{eqclean} also shows that, as $T_e$ increases, the conduction band becomes increasingly populated, making progressively less efficient the phonon decay channel related to e-h formation and leading to an increase of the phonon decay time (Fig.\ref{fig4}b). This leads to a decrease of FWHM(G) for increasing $T_e$ (black solid line in Fig.\ref{fig3}e), which is in contrast with the experimentally observed increase (blue circles in Fig.\ref{fig3}e).

A more realistic description may be obtained by accounting for the effect of $T_e$ on the energy broadening ($\gamma_e$) of the linear dispersion $M_s(z,\epsilon)$, along with the smearing of the Fermi function. $\gamma_{e}(z,T_e)$ can be expressed, to a first approximation, as the sum of three terms\cite{Venezuela2011}:
\begin{equation}
\gamma_e(z,T_e)=\gamma_{ee}(T_e)+\gamma_{ep}(z)+\gamma_{def}(z)
\label{eqgammae}
\end{equation}
where $\gamma_{ee}$, $\gamma_{ep}$ and $\gamma_{def}$ are the e-e, e-ph and defect contributions to $\gamma_e$. The only term that significantly depends on $T_e$ is $\gamma_{ee}$, while the others depend on the energy $z$\cite{Pisana2007,Pinczuk2007, Neumann2015,Bonini2007,Venezuela2011,BaskPRB2009}.

The linear dependence of $\gamma_{ee}$ on T$_e$\cite{Schutt2011} can be estimated from its impact on FWHM(2D). The variation of FWHM(2D) with respect to RT can be written as\cite{Basko2008}:
\begin{equation}
\Delta \textnormal{FWHM(2D)}=4\sqrt{2^{2/3}-1}\frac{1}{2}\frac{\partial \textnormal{POS}(\textnormal{2D})}{\partial (h\nu_{laser})} \gamma_{ee}
\label{eq:gamma2D}
\end{equation}
where $[{\partial \textnormal{POS}(\textnormal{2D})}/{\partial (h\nu_{laser})}]/2 =\frac{1}{c h} v_{ph}/v_F\sim100$cm$^{-1}$/eV \cite{Vidano1981,FerrariNN2013}, i.e. the ratio between the phonon and Fermi velocity, defined as the slope of the phononic (electronic) dispersion at the ph (e) momentum corresponding to a given excitation laser energy $h \nu_{laser}$\cite{FerrariNN2013}. Since the DR process responsible for the 2D peak involves the creation of e-h pairs at energy $\mp h\nu_{laser}/2$, the variation of FWHM(2D) allows us to estimate the variation of $\gamma_e$ at $z=h\nu_{laser}/2\simeq 0.8$eV. Then, $\gamma_{ep}$ and $\gamma_{def}$, both proportional to $z$ ($\gamma_{ep},\gamma_{def}\propto z$), will give an additional constant contribution to FWHM(2D), but not to its variation with $T_e$. Our data support the predicted\cite{Schutt2011} linear increase of $\gamma_{ee}$ with $T_e$, with a dimensionless experimental slope $\alpha_e\simeq 0.73$, Fig.\ref{fig3}c.

In order to compute FWHM(G) from Eq.\ref{EqSE}, we note that the terms $\gamma_{ep}$ and $\gamma_{def}$ are negligible at the relevant low energy $z=\hbar\omega_G/2\sim0.1$eV $\ll  h\nu_{laser}/2$. Hence $\gamma_e(z,T_e)\simeq \gamma_{ee}(T_e)$.

The Dirac cone broadening can now be introduced by accounting for $\gamma_{e}$ in the spectral function of Eq.\ref{ms}:
\begin{equation}
\label{ms2}
M_s(z)=\frac{1}{\pi}\frac{\gamma_e/2}{(z+s\epsilon)^2+(\gamma_e/2)^2}, \quad s=\pm1,
\end{equation}
accordingly, all the processes where the energy difference $|s\epsilon(k)-s'\epsilon(k')+\hbar\omega_0|$ is less than $2\gamma_e$ (which guarantees the overlap between the spectral functions of the quasiparticles) will now contribute in Eq.\ref{EqSE}. Amongst them, those transitions within the same (valence or conduction) band, as shown in Fig.\ref{fig4}d.

The broadened interband contributions still follow, approximately, Eq.\ref{eqclean} (see Fig.\ref{fig4}e). However, the Dirac cone broadening gives additional channels for G phonon annihilation by carrier excitation. In particular, intra-band transitions within the Dirac cone are now progressively enabled for increasing $T_e$, as sketched in Fig.\ref{fig4}d. In Fig.\ref{fig4}e the corresponding contributions to FWHM(G) are shown. Calculations in the weak-coupling limit\cite{Schutt2011} suggest that $\gamma_e(T_e)$ should be suppressed as $z\rightarrow 0$, due to phase-space restriction of the Dirac-cone dispersion. Our results, however, indicate that this effect should appear at an energy scale smaller than $\hbar\omega_G/2$, as the theory captures the main experimental trends, just based on a $z$-independent $\gamma_e(T_e)$.

Critically, the G peak broadening has a different origin from the equilibrium case\cite{Baladin2008}. The absence of anharmonicity would imply a FWHM(G) decrease with temperature due to the e-ph mechanism. However, the Dirac cone broadening reverses this trend into a linewidth broadening above $T_e=1000$K producing, in turn, a dephasing time reduction, corresponding to the experimentally observed FWHM(G) increase. The blueshift of the G peak with temperature is captured by the standard e-ph interaction, beyond possible calibration accuracy. Importantly, the Dirac cone broadening does not significantly affect Pos(G).

In conclusion, we measured the Raman spectrum of SLG with impulsive excitation, in the presence of a distribution of hot charge carriers. Our excitation bandwidth enables us to combine frequency resolution, required to observe the Raman spectra, with short pulse duration, needed to create a significant population of hot carriers. We show that, under these strongly non-equilibrium conditions, the Raman spectrum of graphene cannot be understood based on the standard low fluence picture, and we provide the experimental demonstration of a broadening of the electronic linear dispersion induced by the highly excited carriers. Our results shed light on a novel regime of non-equilibrium Raman response, whereby the e-ph interaction is enhanced. This has implications for the understanding of transient charge carrier mobility under photoexcitation, important to study SLG-based optoelectronic and photonic devices\cite{PhysRevB.93.075410,Mogulkoc201485}, such as broadband light emitters\cite{kim_bright_2015}, transistors and optical gain media\cite{engel_lightmatter_2012}.
\section*{Methods}
\subsection*{Sample preparation and CW Raman characterization}

\begin{figure}
	\centerline{\includegraphics[width=80mm]{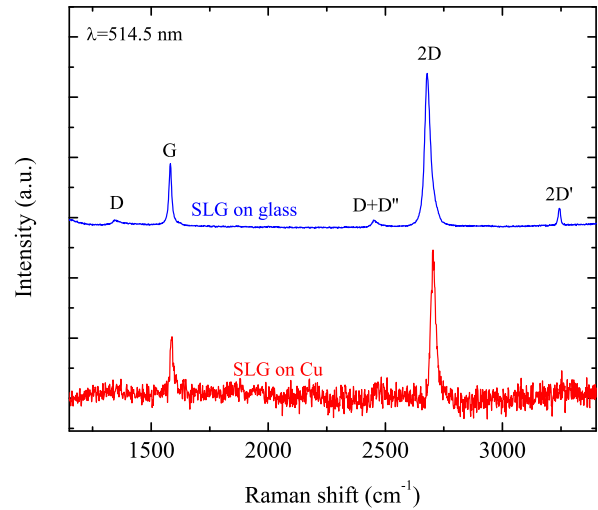}}
	\caption{\textbf{CW Raman spectra of SLG.} Raman response of SLG on Cu (red line), and on glass (blue line) after the transfer from Cu substrate. In the latter case, the substrate spectrum is subtracted.}
	\label{figSRaman}
\end{figure}

SLG is grown on a 35$\mu$m Cu foil, following the process described in Refs.\citenum{BaeNN2010},\citenum{LiS2009}. The substrate is heated to 1000$^{\circ}$C and annealed in hydrogen (H$_2$, 20 sccm) for 30 minutes. Then, 5 sccm of methane (CH$_4$) is let into the chamber for the following 30 minutes so that the growth can take place\cite{BaeNN2010,LiS2009}. The sample is then cooled back to RT in vacuum ($\sim$1 mTorr) and unloaded from the chamber. The sample is characterized by CW Raman spectroscopy using a Renishaw inVia Spectrometer equipped with a 100x objective. The Raman spectrum measured at 514 nm is shown in Fig.\ref{figSRaman} (red curve). This is obtained by removing the non-flat background Cu PL\cite{LagaAPL2013}. The absence of a significant D peak implies negligible defects\cite{FerrariNN2013,FerrariPRL2006,CancNL2011,FerrariPRB2000}. The 2D peak is a single sharp Lorentzian with FWHM(2D)$\sim$23cm$^{-1}$, a signature of SLG\cite{FerrariPRL2006}. Pos(G) is$\sim$1587cm$^{-1}$, with FWHM(G)$\sim$14cm$^{-1}$. Pos(2D) is$\sim$2705cm$^{-1}$, while the 2D to G peak area ratio is $\sim$4.3. SLG is then transferred on glass by a wet method\cite{bonaccorso2012}. Poly-methyl methacrylate (PMMA) is spin coated on the substrate, which is then placed in a solution of ammonium persulfate (APS) and deionized water. Cu is etched\cite{BaeNN2010,bonaccorso2012}, the PMMA membrane with attached SLG is then moved to a beaker with deionized water to remove APS residuals. The membrane is subsequently lifted with the target substrate. After drying, PMMA is removed in acetone leaving SLG on glass. The SLG quality is also monitored after transfer. The Raman spectrum of the substrate shows features in the D and G peak range, convoluted with the spectrum of SLG on glass (blue curve in Fig.\ref{figSRaman}). A point-to-point subtraction is needed to reveal the SLG features. After transfer, the D peak is still negligible, demonstrating that no significant additional defects are induced by the transfer process, and the fitted Raman parameters indicate p doping$\sim$250meV\cite{BaskPRB2009, DasNN2008}.

Before and after the pulsed laser experiment, equilibrium CW measurements are performed at room temperature using a micro-Raman setup (LabRAM Infinity). A different energy and momentum of the D phonon is involved, for a given excitation wavelength, in the S or AS processes, due to the phonon dispersion in the DR mechanism\cite{baranov_1987,thomsen_2000}. Hence, in order to measure the same D phonon in S and AS, different laser excitations (${\nu}_{laser}$) must be used according to ${\nu}_{laser}^{S}={\nu}_{laser}^{AS}+c\textnormal{Pos}(\textnormal{2D})$\cite{FerrariNN2013,tan2002,cancado2002}. Given our pulsed laser wavelength (783nm), the corresponding CW excitation would be$\sim$649.5nm. Hence, we use a 632.8nm He-Ne source, accounting for the small residual wavelength mismatch by scaling the phonon frequency as $\frac{d \textnormal{Pos}(\textnormal{2D})}{d {\nu}_{laser}}=0.0132/c$\cite{FerrariNN2013}
\subsection*{Pulsed Raman measurements}

The ps-Raman apparatus is based on a mode-locked Er:fiber amplified laser at$\sim1550$nm, producing 90fs pulses at a repetition rate RR=40MHz. Using second-harmonic generation in a 1cm Periodically Poled Lithium Niobate crystal\cite{Marangoni2007}, we obtain 3ps pulses at 783nm with a$\sim9.5$cm$^{-1}$ bandwidth. The beam is focused on SLG through a slightly underfilled 20X objective (NA$=0.4$), resulting in a focal diameter $D=5.7\mu$m. Back-scattered light is collected by the same objective, separated with a dichroic filter from the incident beam and sent to a spectrometer (with a resolution of 0.028 nm/pixel corresponding to 1.2cm$^{-1}$). The overall IRF, therefore, is dominated by the additional contribution induced by the finite excitation pulse bandwidth. Hence, in order to extract the FWHM of the Raman peaks, our data are fitted convolving a Lorentzian with the spectral profile of the laser excitation.

When using ultrafast pulses, a non-linear PL is seen in SLG\cite{Lui2010}. Such an effect is particularly intense for the S spectral range\cite{PhysRevB.82.081408,Stohr2010}. The S signal in Fig.\ref{fig1}c is obtained as the difference spectrum of two measurements with excitation frequencies slightly offset by$\sim$130cm$^{-1}$, resulting in PL suppression. The background subtraction requires in this case a wider spectral range, at the expenses of spectrometer resolution which is reduced to 0.13 nm/pixel corresponding to$\sim$6cm$^{-1}$, as additional contribution to the IRF. Although this procedure allows to isolate the S Raman peaks, the resulting noise level is worse than for AS. For this reason we mostly focus on the AS features.

\begin{figure}[h!]
	\centerline{\includegraphics[width=90mm]{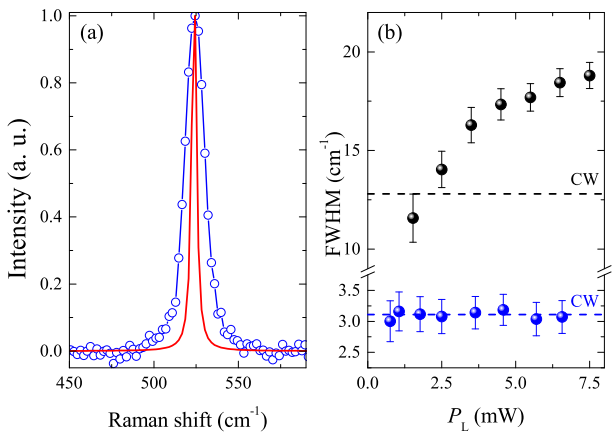}}
	\caption{\textbf{Raman response of Si for pulsed laser excitation.}(a)Raman spectrum of Si measured for ultrafast laser excitation and 6.6mW average power. (blue line) Lorentzian fit. (red line) laser-bandwidth deconvoluted spectrum. (b) FWHM(Si) as a function of $P_L$ (blue symbols) does not show any deviation from the CW FWHM(Si) (dashed blue line). FWHM(G) under the same excitation conditions (black symbols) deviates from the CW regime (dashed black line).}
	\label{figS1}
\end{figure}

To verify that the observed peaks broadening is not limited by our IRF, we perform the same experiment on a Si substrate (\ref{figS1}a). For this we retrieve, after deconvolution of the IRF, the same Raman linewidth measured in the CW excitation regime (Fig.\ref{figS1}a). The FWHM of the Si optical phonon is independent of $P_L$, in contrast with the well-defined dependence on $P_L$ observed in SLG, Fig.\ref{figS1}b.
\subsection*{Estimate of the local equilibrium temperature $T_{eq}$}
Photoexcitation of SLG induces an excess of energy in the form of heat Q per unit area, that can be expressed as:
\begin{equation}
Q\sim\frac{P_L}{RR}\frac{A}{\pi W^2}
\label{Q1}
\end{equation}
where $A=2.3$\% is the SLG absorption, approximated to the undoped case\cite{Nair2008}, $W\sim2.8\mu$m is the waist of focused beam and $RR=40$MHz is the repetition rate of the excitation laser.
The induced $T_{eq}$ can be derived based on two assumptions: (i) in our ps time scale the energy absorbed in the focal region does not diffuse laterally, (ii) the energy is equally distributed on each degree of freedom (electrons, optical and acoustic ph). Then, $Q$ can be described as:
\begin{equation}
Q=\int_{RT}^{T_{eq}}C(T)\,dT
\label{Q2}
\end{equation}
where $C(T)$ is the SLG T-dependent specific heat. In the $300-700$K range, $C(T)$ can be described as\cite{Pop2012}: $C(T)=aT+b$, where $a=1.35\cdot10^{-6}$J/(K$^2\cdot$m$^2$) and $b=1.35\cdot10^{-4}$J/(K$\cdot$m$^2$). Therefore, considering Eqs.\ref{Q1},\ref{Q2}, for $P_L=P^{max}=13.5$mW, we get $T_{eq}\sim680$K, well below the corresponding $T_e$, indicating an out-of-equilibrium condition ($T_l<T_{eq}<T_{e}$). Any contributions from the substrate and taking into account for the heat profile would contribute in reducing even further $T_l$ estimation.
\subsection*{Estimate of Pos(2D) as a function of $T_e$}

\begin{figure}
	\centerline{\includegraphics[width=90mm]{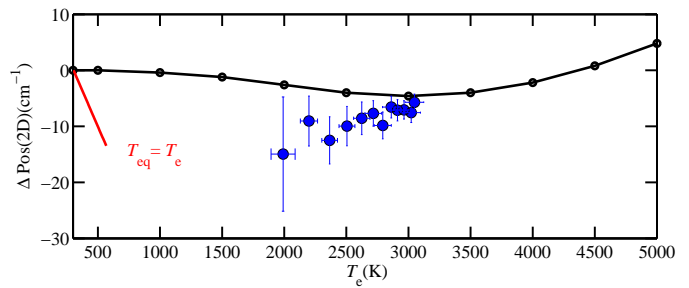}}
	\caption{\textbf{Temperature dependence of Pos(2D).} Pos(2D), relative to the RT CW measurement, as a function of $T_e$. Black line: DFPT; Blue circles: experimental data with pulsed excitation. Red line: T-dependent CW measurement in thermal equilibrium ($T_e=T_l=T_{eq}$) from Ref.\cite{Nef2014}.}
	\label{figS2}
\end{figure}
We perform calculations within the Local Density Approximation in DFPT\cite{Gironcoli1995,Giannozzi2009}. We use the experimental lattice parameter
2.46\AA\cite{wang_crystal_2012} and plane waves (45Ry cutoff), within a norm-conserving pseudopotential approach\cite{Giannozzi2009}. The electronic levels are occupied with a finite fictitious $T_e$ with a Fermi Dirac distribution, and we sample a Brillouin Zone with a 160x160x1 mesh. This does not take into account anharmonic effects, assuming $T_l=300$K. Fig.\ref{figS2} shows a weak $\Delta Pos(2D)$($\sim 5$cm$^{-1}$) in the range $T_e=300-3000$K. In equilibrium, $T_l=T_e$ would induce a non-negligible anharmonicity\cite{Nef2014}, which would lead to a Pos(2D) softening: $\Delta Pos(2D)/\Delta T_{eq}=-0.05$cm$^{-1}/$K. The weak dependence $\Delta Pos(2D)(P_L)$ (blue circles in Fig.\ref{figS2}) rules out a dominant anharmonicity contribution and, consequently, $T_l=T_e$. The minor disagreement with DFPT suggests a $T_l$ slightly larger than RT, but definitely smaller than $T_{eq}$.
\section*{Acknowledgments}
We acknowledge funding from the Graphene Flagship, ERC Grant Hetero2D,  EPSRC Grants EP/K01711X/1, EP/K017144/1, EP/N010345/1,  EP/L016087/1 and MAECI under the Italia-India collaborative project SuperTop-PGR04879.
\section*{Author Contributions}
TS led the research project, conceived with GC, FM and ACF.
CFe, AV and TS designed and built the pulsed Raman setup.
CFe and AV performed the out of equilibrium Raman experiments, with contribution from MM.
CFa, PP, DDF, US, AKO and DY performed the equilibrium CW Raman experiment.
LB and FM developed the modelling and carried out the numerical simulations, with contribution from AV
DDF, US, AKO and DY prepared and characterised the sample.
CFe, AV, LB, GC, FM, ACF and TS interpreted the data and the simulations and wrote the manuscript.
\bibliographystyle{naturemag}

\begin{thebibliography}{10}
	\expandafter\ifx\csname url\endcsname\relax
	\def\url#1{\texttt{#1}}\fi
	\expandafter\ifx\csname urlprefix\endcsname\relax\def\urlprefix{URL }\fi
	\providecommand{\bibinfo}[2]{#2}
	\providecommand{\eprint}[2][]{\url{#2}}
	
	\bibitem{Bonaccorso2010}
	\bibinfo{author}{Bonaccorso, F.}, \bibinfo{author}{Sun, Z.},
	\bibinfo{author}{Hasan, T.} \& \bibinfo{author}{Ferrari, A.~C.}
	\newblock \bibinfo{title}{Graphene photonics and optoelectronics}.
	\newblock \emph{\bibinfo{journal}{Nat. Photonics}} \textbf{\bibinfo{volume}{4}},
	\bibinfo{pages}{611--622} (\bibinfo{year}{2010}).
	
	\bibitem{Nanoscale2015}
	\bibinfo{author}{Ferrari, A.~C.} \emph{et~al.}
	\newblock \bibinfo{title}{Science and technology roadmap for graphene{,}
		related two-dimensional crystals{,} and hybrid systems}.
	\newblock \emph{\bibinfo{journal}{Nanoscale}} \textbf{\bibinfo{volume}{7}},
	\bibinfo{pages}{4598--4810} (\bibinfo{year}{2015}).
	
	\bibitem{Koppens2014}
	\bibinfo{author}{Koppens, F. H.~L.} \emph{et~al.}
	\newblock \bibinfo{title}{Photodetectors based on graphene, other
		two-dimensional materials and hybrid systems}.
	\newblock \emph{\bibinfo{journal}{Nat. Nanotech.}}
	\textbf{\bibinfo{volume}{9}}, \bibinfo{pages}{780--793}
	(\bibinfo{year}{2014}).
	
	\bibitem{Brida2013}
	\bibinfo{author}{Brida, D.} \emph{et~al.}
	\newblock \bibinfo{title}{Ultrafast collinear scattering and carrier
		multiplication in graphene}.
	\newblock \emph{\bibinfo{journal}{Nat. Commun.}} \textbf{\bibinfo{volume}{4}}
	(\bibinfo{year}{2013}).
	
	\bibitem{Tomadin2013}
	\bibinfo{author}{Tomadin, A.}, \bibinfo{author}{Brida, D.},
	\bibinfo{author}{Cerullo, G.}, \bibinfo{author}{Ferrari, A.~C.} \&
	\bibinfo{author}{Polini, M.}
	\newblock \bibinfo{title}{Nonequilibrium dynamics of photoexcited electrons in
		graphene: Collinear scattering, auger processes, and the impact of
		screening}.
	\newblock \emph{\bibinfo{journal}{Phys. Rev. B}} \textbf{\bibinfo{volume}{88}},
	\bibinfo{pages}{035430} (\bibinfo{year}{2013}).
	
	\bibitem{Lazzeri2005}
	\bibinfo{author}{Lazzeri, M.}, \bibinfo{author}{Piscanec, S.},
	\bibinfo{author}{Mauri, F.}, \bibinfo{author}{Ferrari, A.~C.} \&
	\bibinfo{author}{Robertson, J.}
	\newblock \bibinfo{title}{Electron transport and hot phonons in carbon
		nanotubes}.
	\newblock \emph{\bibinfo{journal}{Phys. Rev. Lett.}}
	\textbf{\bibinfo{volume}{95}}, \bibinfo{pages}{236802}
	(\bibinfo{year}{2005}).
	
	\bibitem{Butscher2007}
	\bibinfo{author}{Butscher, S.}, \bibinfo{author}{Milde, F.},
	\bibinfo{author}{Hirtschulz, M.}, \bibinfo{author}{Mali\`c, E.} \&
	\bibinfo{author}{Knorr, A.}
	\newblock \bibinfo{title}{Hot electron relaxation and phonon dynamics in
		graphene}.
	\newblock \emph{\bibinfo{journal}{Appl. Phys. Lett.}}
	\textbf{\bibinfo{volume}{91}}, \bibinfo{pages}{203103}
	(\bibinfo{year}{2007}).
	
	\bibitem{Lui2010}
	\bibinfo{author}{Lui, C.~H.}, \bibinfo{author}{Mak, K.~F.},
	\bibinfo{author}{Shan, J.} \& \bibinfo{author}{Heinz, T.~F.}
	\newblock \bibinfo{title}{Ultrafast photoluminescence from graphene}.
	\newblock \emph{\bibinfo{journal}{Phys. Rev. Lett.}}
	\textbf{\bibinfo{volume}{105}}, \bibinfo{pages}{127404}
	(\bibinfo{year}{2010}).
	
	\bibitem{Wu2012}
	\bibinfo{author}{Wu, S.} \emph{et~al.}
	\newblock \bibinfo{title}{Hot phonon dynamics in graphene}.
	\newblock \emph{\bibinfo{journal}{Nano Lett.}} \textbf{\bibinfo{volume}{12}},
	\bibinfo{pages}{5495--5499} (\bibinfo{year}{2012}).
	
	\bibitem{Bonini2007}
	\bibinfo{author}{Bonini, N.}, \bibinfo{author}{Lazzeri, M.},
	\bibinfo{author}{Marzari, N.} \& \bibinfo{author}{Mauri, F.}
	\newblock \bibinfo{title}{Phonon anharmonicities in graphite and graphene}.
	\newblock \emph{\bibinfo{journal}{Phys. Rev. Lett.}}
	\textbf{\bibinfo{volume}{99}}, \bibinfo{pages}{176802}
	(\bibinfo{year}{2007}).
	
	\bibitem{FerrRob2004}
	\bibinfo{author}{Ferrari, A.~C.} \& \bibinfo{author}{Robertson, J.~e.}
	\newblock \bibinfo{title}{Raman spectroscopy in carbons: from nanotubes to
		diamond, theme issue}.
	\newblock \emph{\bibinfo{journal}{Phil. Trans. R. Soc. Lond. A}}
	\textbf{\bibinfo{volume}{362}}, \bibinfo{pages}{2267--2565}
	(\bibinfo{year}{2004}).
	
	\bibitem{FerrariPRL2006}
	\bibinfo{author}{Ferrari, A.~C.} \emph{et~al.}
	\newblock \bibinfo{title}{Raman spectrum of graphene and graphene layers}.
	\newblock \emph{\bibinfo{journal}{Phys. Rev. Lett.}}
	\textbf{\bibinfo{volume}{97}}, \bibinfo{pages}{187401}
	(\bibinfo{year}{2006}).
	
	\bibitem{FerrariNN2013}
	\bibinfo{author}{Ferrari, A.~C.} \& \bibinfo{author}{Basko, D.~M.}
	\newblock \bibinfo{title}{Raman spectroscopy as a versatile tool for studying
		the properties of graphene}.
	\newblock \emph{\bibinfo{journal}{Nat. Nanotech.}}
	\textbf{\bibinfo{volume}{8}}, \bibinfo{pages}{235--246}
	(\bibinfo{year}{2013}).
	
	\bibitem{Malard2009}
	\bibinfo{author}{Malard, L.}, \bibinfo{author}{Pimenta, M.},
	\bibinfo{author}{Dresselhaus, G.} \& \bibinfo{author}{Dresselhaus, M.}
	\newblock \bibinfo{title}{Raman spectroscopy in graphene}.
	\newblock \emph{\bibinfo{journal}{Phys. Rep.}}
	\textbf{\bibinfo{volume}{473}}, \bibinfo{pages}{51 -- 87}
	(\bibinfo{year}{2009}).
	
	\bibitem{Froehlicher2015}
	\bibinfo{author}{Froehlicher, G.} \& \bibinfo{author}{Berciaud, S.}
	\newblock \bibinfo{title}{Raman spectroscopy of electrochemically gated
		graphene transistors: Geometrical capacitance, electron-phonon,
		electron-electron, and electron-defect scattering}.
	\newblock \emph{\bibinfo{journal}{Phys. Rev. B}} \textbf{\bibinfo{volume}{91}},
	\bibinfo{pages}{205413} (\bibinfo{year}{2015}).
	
	\bibitem{Tan2012}
	\bibinfo{author}{Tan, P.~H.} \emph{et~al.}
	\newblock \bibinfo{title}{The shear mode of multilayer graphene}.
	\newblock \emph{\bibinfo{journal}{Nat. Mater.}} \textbf{\bibinfo{volume}{11}},
	\bibinfo{pages}{294--300} (\bibinfo{year}{2012}).
	
	\bibitem{Sato2011}
	\bibinfo{author}{Sato, K.} \emph{et~al.}
	\newblock \bibinfo{title}{Raman spectra of out-of-plane phonons in bilayer
		graphene}.
	\newblock \emph{\bibinfo{journal}{Phys. Rev. B}} \textbf{\bibinfo{volume}{84}},
	\bibinfo{pages}{035419} (\bibinfo{year}{2011}).
	
	\bibitem{Lui2012}
	\bibinfo{author}{Lui, C.~H.} \emph{et~al.}
	\newblock \bibinfo{title}{Observation of layer-breathing mode vibrations in
		few-layer graphene through combination raman scattering}.
	\newblock \emph{\bibinfo{journal}{Nano Lett.}} \textbf{\bibinfo{volume}{12}},
	\bibinfo{pages}{5539--5544} (\bibinfo{year}{2012}).
	
	\bibitem{Tuinstra1970}
	\bibinfo{author}{Tuinstra, F.} \& \bibinfo{author}{Koenig, J.~L.}
	\newblock \bibinfo{title}{Raman spectrum of graphite}.
	\newblock \emph{\bibinfo{journal}{J. Chem. Phys.}}
	\textbf{\bibinfo{volume}{53}}, \bibinfo{pages}{1126--1130}
	(\bibinfo{year}{1970}).
	
	\bibitem{Thomsen2000}
	\bibinfo{author}{Thomsen, C.} \& \bibinfo{author}{Reich, S.}
	\newblock \bibinfo{title}{Double resonant raman scattering in graphite}.
	\newblock \emph{\bibinfo{journal}{Phys. Rev. Lett.}}
	\textbf{\bibinfo{volume}{85}}, \bibinfo{pages}{5214--5217}
	(\bibinfo{year}{2000}).
	
	\bibitem{FerrariPRB2000}
	\bibinfo{author}{Ferrari, A.~C.} \& \bibinfo{author}{Robertson, J.}
	\newblock \bibinfo{title}{Interpretation of raman spectra of disordered and
		amorphous carbon}.
	\newblock \emph{\bibinfo{journal}{Phys. Rev. B}} \textbf{\bibinfo{volume}{61}},
	\bibinfo{pages}{14095--14107} (\bibinfo{year}{2000}).
	
	\bibitem{Piscanec2004}
	\bibinfo{author}{Piscanec, S.}, \bibinfo{author}{Lazzeri, M.},
	\bibinfo{author}{Mauri, F.}, \bibinfo{author}{Ferrari, A.~C.} \&
	\bibinfo{author}{Robertson, J.}
	\newblock \bibinfo{title}{Kohn anomalies and electron-phonon interactions in
		graphite}.
	\newblock \emph{\bibinfo{journal}{Phys. Rev. Lett.}}
	\textbf{\bibinfo{volume}{93}}, \bibinfo{pages}{185503}
	(\bibinfo{year}{2004}).
	
	\bibitem{Yan2009}
	\bibinfo{author}{Yan, H.} \emph{et~al.}
	\newblock \bibinfo{title}{Time-resolved raman spectroscopy of optical phonons
		in graphite: Phonon anharmonic coupling and anomalous stiffening}.
	\newblock \emph{\bibinfo{journal}{Phys. Rev. B}} \textbf{\bibinfo{volume}{80}},
	\bibinfo{pages}{121403} (\bibinfo{year}{2009}).
	
	\bibitem{Breusing2011}
	\bibinfo{author}{Breusing, M.} \emph{et~al.}
	\newblock \bibinfo{title}{Ultrafast nonequilibrium carrier dynamics in a single
		graphene layer}.
	\newblock \emph{\bibinfo{journal}{Phys. Rev. B}} \textbf{\bibinfo{volume}{83}},
	\bibinfo{pages}{153410} (\bibinfo{year}{2011}).
	
	\bibitem{Chae2010}
	\bibinfo{author}{Chae, D.-H.}, \bibinfo{author}{Krauss, B.},
	\bibinfo{author}{von Klitzing, K.} \& \bibinfo{author}{Smet, J.~H.}
	\newblock \bibinfo{title}{Hot phonons in an electrically biased graphene
		constriction}.
	\newblock \emph{\bibinfo{journal}{Nano Letters}} \textbf{\bibinfo{volume}{10}},
	\bibinfo{pages}{466--471} (\bibinfo{year}{2010}).
	
	\bibitem{PhysRevLett.104.227401}
	\bibinfo{author}{Berciaud, S.} \emph{et~al.}
	\newblock \bibinfo{title}{Electron and optical phonon temperatures in
		electrically biased graphene}.
	\newblock \emph{\bibinfo{journal}{Phys. Rev. Lett.}}
	\textbf{\bibinfo{volume}{104}}, \bibinfo{pages}{227401}
	(\bibinfo{year}{2010}).
	\newblock
	
	\bibitem{PhysRevB.93.075410}
	\bibinfo{author}{McKitterick, C.~B.}, \bibinfo{author}{Prober, D.~E.} \&
	\bibinfo{author}{Rooks, M.~J.}
	\newblock \bibinfo{title}{Electron-phonon cooling in large monolayer graphene
		devices}.
	\newblock \emph{\bibinfo{journal}{Phys. Rev. B}} \textbf{\bibinfo{volume}{93}},
	\bibinfo{pages}{075410} (\bibinfo{year}{2016}).
	
	\bibitem{Mogulkoc201485}
	\bibinfo{author}{Mogulkoc, A.} \emph{et~al.}
	\newblock \bibinfo{title}{The role of electron--phonon interaction on the
		transport properties of graphene based nano-devices}.
	\newblock \emph{\bibinfo{journal}{Physica B}}
	\textbf{\bibinfo{volume}{446}}, \bibinfo{pages}{85 -- 91}
	(\bibinfo{year}{2014}).
	
	\bibitem{kim_bright_2015}
	\bibinfo{author}{Kim, Y.~D.} \emph{et~al.}
	\newblock \bibinfo{title}{Bright visible light emission from graphene}.
	\newblock \emph{\bibinfo{journal}{Nat. Nano.}} \textbf{\bibinfo{volume}{10}},
	\bibinfo{pages}{676--681} (\bibinfo{year}{2015}).
	
	\bibitem{sun2010}
	\bibinfo{author}{Sun, Z.} \emph{et~al.}
	\newblock \bibinfo{title}{Graphene mode-locked ultrafast laser}.
	\newblock \emph{\bibinfo{journal}{ACS Nano}} \textbf{\bibinfo{volume}{4}},
	\bibinfo{pages}{803--810} (\bibinfo{year}{2010}).
	
	\bibitem{Liu2011}
	\bibinfo{author}{Liu, M.} \emph{et~al.}
	\newblock \bibinfo{title}{A graphene-based broadband optical modulator}.
	\newblock \emph{\bibinfo{journal}{Nature}} \textbf{\bibinfo{volume}{474}},
	\bibinfo{pages}{64--67} (\bibinfo{year}{2011}).
	
	\bibitem{papoulis_1962}
	\bibinfo{author}{Papoulis, A.}
	\newblock \emph{\bibinfo{title}{The {Fourier} {Integral} and {Its}
			{Applications}}} (\bibinfo{publisher}{Mcgraw-Hill College},
	\bibinfo{address}{New York}, \bibinfo{year}{1962}).
	
	\bibitem{freitag_2010}
	\bibinfo{author}{Freitag, M.}, \bibinfo{author}{Chiu, H.-Y.},
	\bibinfo{author}{Steiner, M.}, \bibinfo{author}{Perebeinos, V.} \&
	\bibinfo{author}{Avouris, P.}
	\newblock \bibinfo{title}{Thermal infrared emission from biased graphene}.
	\newblock \emph{\bibinfo{journal}{Nat. Nanotech.}} \textbf{\bibinfo{volume}{5}},
	\bibinfo{pages}{497--501} (\bibinfo{year}{2010}).
	
	\bibitem{Stohr2010}
	\bibinfo{author}{St\"ohr, R.~J.}, \bibinfo{author}{Kolesov, R.},
	\bibinfo{author}{Pflaum, J.} \& \bibinfo{author}{Wrachtrup, J.}
	\newblock \bibinfo{title}{Fluorescence of laser-created electron-hole plasma in
		graphene}.
	\newblock \emph{\bibinfo{journal}{Phys. Rev. B}} \textbf{\bibinfo{volume}{82}},
	\bibinfo{pages}{121408} (\bibinfo{year}{2010}).
	
	\bibitem{Gokus2009}
	\bibinfo{author}{Gokus, T.} \emph{et~al.}
	\newblock \bibinfo{title}{Making graphene luminescent by oxygen plasma
		treatment}.
	\newblock \emph{\bibinfo{journal}{ACS Nano}} \textbf{\bibinfo{volume}{3}},
	\bibinfo{pages}{3963--3968} (\bibinfo{year}{2009}).
	
	\bibitem{callen_1985}
	\bibinfo{author}{Callen, H.~B.}
	\newblock \emph{\bibinfo{title}{Thermodynamics and an {Introduction} to
			{Thermostatistics}}} (\bibinfo{publisher}{John Wiley \& Sons Inc},
	\bibinfo{address}{New York}, \bibinfo{year}{1985}).
	
	\bibitem{princeton} www.princetoninstruments.com/calculators/grating-dispersion.cfm
	
	\bibitem{princeton2} www.princetoninstruments.com/userfiles/files/\\assetLibrary/Datasheets/Princeton-Instruments-PIXIS-100-rev-5-1-10-22-14.pdf
	
	\bibitem{Schomacker1986}
	\bibinfo{author}{Schomacker, K.~T.} \& \bibinfo{author}{Champion, P.~M.}
	\newblock \bibinfo{title}{Investigations of spectral broadening mechanisms in biomolecules: Cytochrome--c}.
	\newblock \emph{\bibinfo{journal}{J. Chem. Phys.}}
	\textbf{\bibinfo{volume}{84}}, \bibinfo{pages}{5314--5325}
	(\bibinfo{year}{1986}).
	
	\bibitem{FilhoPRB2001}
	\bibinfo{author}{Souza~Filho, A.~G.} \emph{et~al.}
	\newblock \bibinfo{title}{Electronic transition energy ${E}_{\mathrm{ii}}$ for
		an isolated $(n,m)$ single-wall carbon nanotube obtained by
		anti-stokes/stokes resonant raman intensity ratio}.
	\newblock \emph{\bibinfo{journal}{Phys. Rev. B}} \textbf{\bibinfo{volume}{63}},
	\bibinfo{pages}{241404} (\bibinfo{year}{2001}).
	
	\bibitem{goldstein2016raman}
	\bibinfo{author}{Goldstein, T.} \emph{et~al.}
	\newblock \bibinfo{title}{Raman scattering and anomalous stokes--anti-stokes
		ratio in MoTe$_2$ atomic layers}.
	\newblock \emph{\bibinfo{journal}{Sci. Rep.}}
	\textbf{\bibinfo{volume}{6}}, \bibinfo{pages}{28024} (\bibinfo{year}{2016}).
	
	\bibitem{ParraMurilloPRB2016}
	\bibinfo{author}{Parra-Murillo, C.~A.}, \bibinfo{author}{Santos, M.~F.},
	\bibinfo{author}{Monken, C.~H.} \& \bibinfo{author}{Jorio, A.}
	\newblock \bibinfo{title}{Stokes--anti-Stokes correlation in the
		inelastic scattering of light by matter and generalization of the
		bose-einstein population function}.
	\newblock \emph{\bibinfo{journal}{Phys. Rev. B}} \textbf{\bibinfo{volume}{93}},
	\bibinfo{pages}{125141} (\bibinfo{year}{2016}).
	
	\bibitem{Apostolov}
	\bibinfo{author}{Apostolov, A.~T.}, \bibinfo{author}{Apostolova, I.~N.} \&
	\bibinfo{author}{Wesselinowa, J.~M.}
	\newblock \bibinfo{title}{Temperature and layer number dependence of the g and
		2d phonon energy and damping in graphene}.
	\newblock \emph{\bibinfo{journal}{J. Phys. Condens. Matter}}
	\textbf{\bibinfo{volume}{24}}, \bibinfo{pages}{235401}
	(\bibinfo{year}{2012}).
	
	\bibitem{Basko2008}
	\bibinfo{author}{Basko, D.~M.}
	\newblock \bibinfo{title}{Theory of resonant multiphonon raman scattering in
		graphene}.
	\newblock \emph{\bibinfo{journal}{Phys. Rev. B}} \textbf{\bibinfo{volume}{78}},
	\bibinfo{pages}{125418} (\bibinfo{year}{2008}).
	
	\bibitem{LazzeriPRL06}
	\bibinfo{author}{Lazzeri, M.} \& \bibinfo{author}{Mauri, F.}
	\newblock \bibinfo{title}{Nonadiabatic kohn anomaly in a doped graphene
		monolayer}.
	\newblock \emph{\bibinfo{journal}{Phys. Rev. Lett.}}
	\textbf{\bibinfo{volume}{97}}, \bibinfo{pages}{266407}
	(\bibinfo{year}{2006}).
	
	\bibitem{Pisana2007}
	\bibinfo{author}{Pisana, S.} \emph{et~al.}
	\newblock \bibinfo{title}{Breakdown of the adiabatic born-oppenheimer
		approximation in graphene}.
	\newblock \emph{\bibinfo{journal}{Nat. Mater.}} \textbf{\bibinfo{volume}{6}},
	\bibinfo{pages}{198--201} (\bibinfo{year}{2007}).
	
	\bibitem{Ferrari200747}
	\bibinfo{author}{Ferrari, A.~C.}
	\newblock \bibinfo{title}{Raman spectroscopy of graphene and graphite:
		Disorder, electron--phonon coupling, doping and nonadiabatic effects}.
	\newblock \emph{\bibinfo{journal}{Solid State Commun.}}
	\textbf{\bibinfo{volume}{143}}, \bibinfo{pages}{47 -- 57}
	(\bibinfo{year}{2007}).
	
	\bibitem{Ando2006}
	\bibinfo{author}{Ando, T.}
	\newblock \bibinfo{title}{Anomaly of optical phonon in monolayer graphene}.
	\newblock \emph{\bibinfo{journal}{J. Phys. Soc. Jpn.}}
	\textbf{\bibinfo{volume}{75}}, \bibinfo{pages}{124701}
	(\bibinfo{year}{2006}).
	
	\bibitem{Venezuela2011}
	\bibinfo{author}{Venezuela, P.}, \bibinfo{author}{Lazzeri, M.} \&
	\bibinfo{author}{Mauri, F.}
	\newblock \bibinfo{title}{Theory of double-resonant raman spectra in graphene:
		Intensity and line shape of defect-induced and two-phonon bands}.
	\newblock \emph{\bibinfo{journal}{Phys. Rev. B}} \textbf{\bibinfo{volume}{84}},
	\bibinfo{pages}{035433} (\bibinfo{year}{2011}).
	
	\bibitem{Pinczuk2007}
	\bibinfo{author}{Yan, J.}, \bibinfo{author}{Zhang, Y.}, \bibinfo{author}{Kim,
		P.} \& \bibinfo{author}{Pinczuk, A.}
	\newblock \bibinfo{title}{Electric field effect tuning of electron-phonon
		coupling in graphene}.
	\newblock \emph{\bibinfo{journal}{Phys. Rev. Lett.}}
	\textbf{\bibinfo{volume}{98}}, \bibinfo{pages}{166802}
	(\bibinfo{year}{2007}).
	
	\bibitem{Neumann2015}
	\bibinfo{author}{Neumann, C.} \emph{et~al.}
	\newblock \bibinfo{title}{Raman spectroscopy as probe of nanometre-scale strain
		variations in graphene}.
	\newblock \emph{\bibinfo{journal}{Nat. Commun.}} \textbf{\bibinfo{volume}{6}},
	\bibinfo{pages}{8429} (\bibinfo{year}{2015}).
	
	\bibitem{BaskPRB2009}
	\bibinfo{author}{Basko, D.~M.}, \bibinfo{author}{Piscanec, S.} \&
	\bibinfo{author}{Ferrari, A.~C.}
	\newblock \bibinfo{title}{Electron-electron interactions and doping dependence
		of the two-phonon raman intensity in graphene}.
	\newblock \emph{\bibinfo{journal}{Phys. Rev. B}} \textbf{\bibinfo{volume}{80}},
	\bibinfo{pages}{165413} (\bibinfo{year}{2009}).
	
	\bibitem{Schutt2011}
	\bibinfo{author}{Sch\"utt, M.}, \bibinfo{author}{Ostrovsky, P.~M.},
	\bibinfo{author}{Gornyi, I.~V.} \& \bibinfo{author}{Mirlin, A.~D.}
	\newblock \bibinfo{title}{Coulomb interaction in graphene: Relaxation rates and
		transport}.
	\newblock \emph{\bibinfo{journal}{Phys. Rev. B}} \textbf{\bibinfo{volume}{83}},
	\bibinfo{pages}{155441} (\bibinfo{year}{2011}).
	
	\bibitem{Vidano1981}
	\bibinfo{author}{Vidano, R.}, \bibinfo{author}{Fischbach, D.},
	\bibinfo{author}{Willis, L.} \& \bibinfo{author}{Loehr, T.}
	\newblock \bibinfo{title}{Observation of raman band shifting with excitation
		wavelength for carbons and graphites}.
	\newblock \emph{\bibinfo{journal}{Solid State Commun.}}
	\textbf{\bibinfo{volume}{39}}, \bibinfo{pages}{341 -- 344}
	(\bibinfo{year}{1981}).
	
	\bibitem{Baladin2008}
	\bibinfo{author}{Balandin, A.~A.} \emph{et~al.}
	\newblock \bibinfo{title}{Superior thermal conductivity of single-layer
		graphene}.
	\newblock \emph{\bibinfo{journal}{Nano Lett.}} \textbf{\bibinfo{volume}{8}},
	\bibinfo{pages}{902--907} (\bibinfo{year}{2008}).
	
	
	\bibitem{engel_lightmatter_2012}
	\bibinfo{author}{Engel, M.} \emph{et~al.}
	\newblock \bibinfo{title}{Light--matter interaction in a
		microcavity-controlled graphene transistor}.
	\newblock \emph{\bibinfo{journal}{Nat. Commun.}} \textbf{\bibinfo{volume}{3}},
	\bibinfo{pages}{906} (\bibinfo{year}{2012}).
	
	\bibitem{BaeNN2010}
	\bibinfo{author}{Bae, S.} \emph{et~al.}
	\newblock \bibinfo{title}{Roll-to-roll production of 30-inch graphene films for
		transparent electrodes}.
	\newblock \emph{\bibinfo{journal}{Nat. Nanotech.}}
	\textbf{\bibinfo{volume}{5}}, \bibinfo{pages}{574--578}
	(\bibinfo{year}{2010}).
	
	\bibitem{LiS2009}
	\bibinfo{author}{Li, X.} \emph{et~al.}
	\newblock \bibinfo{title}{Large-area synthesis of high-quality and uniform
		graphene films on copper foils}.
	\newblock \emph{\bibinfo{journal}{Science}} \textbf{\bibinfo{volume}{324}},
	\bibinfo{pages}{1312--1314} (\bibinfo{year}{2009}).
	
	\bibitem{LagaAPL2013}
	\bibinfo{author}{Lagatsky, A.~A.} \emph{et~al.}
	\newblock \bibinfo{title}{2$\mu$m solid-state laser mode-locked by single-layer graphene}.
	\newblock \emph{\bibinfo{journal}{Appl. Phys. Lett.}}
	\textbf{\bibinfo{volume}{102}}, \bibinfo{pages}{013113}
	(\bibinfo{year}{2013}).
	
	\bibitem{CancNL2011}
	\bibinfo{author}{Can\c{c}ado, L.~G.} \emph{et~al.}
	\newblock \bibinfo{title}{Quantifying defects in graphene via raman
		spectroscopy at different excitation energies}.
	\newblock \emph{\bibinfo{journal}{Nano Lett.}} \textbf{\bibinfo{volume}{11}},
	\bibinfo{pages}{3190--3196} (\bibinfo{year}{2011}).
	
	\bibitem{bonaccorso2012}
	\bibinfo{author}{Bonaccorso, F.} \emph{et~al.}
	\newblock \bibinfo{title}{Production and processing of graphene and 2d
		crystals}.
	\newblock \emph{\bibinfo{journal}{Mater. Today}} \textbf{\bibinfo{volume}{15}},
	\bibinfo{pages}{564 -- 589} (\bibinfo{year}{2012}).
	
	\bibitem{DasNN2008}
	\bibinfo{author}{Das, A.} \emph{et~al.}
	\newblock \bibinfo{title}{Monitoring dopants by raman scattering in an
		electrochemically top-gated graphene transistor}.
	\newblock \emph{\bibinfo{journal}{Nat. Nanotech.}}
	\textbf{\bibinfo{volume}{3}}, \bibinfo{pages}{210--215}
	(\bibinfo{year}{2008}).
	
	\bibitem{baranov_1987}
	\bibinfo{author}{Baranov, A.~V.}, \bibinfo{author}{Bekhterev, A.~N.},
	\bibinfo{author}{Bobovich, Y.~S.} \& \bibinfo{author}{Petrov, V.~I.}
	\newblock \bibinfo{title}{Interpretation of certain characteristics in raman
		spectra of graphite and glassy carbon}.
	\newblock \emph{\bibinfo{journal}{Opt. Spectrosc.}}
	\textbf{\bibinfo{volume}{62}} (\bibinfo{year}{1987}).
	
	\bibitem{thomsen_2000}
	\bibinfo{author}{Thomsen, C.} \& \bibinfo{author}{Reich, S.}
	\newblock \bibinfo{title}{Double {Resonant} {Raman} {Scattering} in
		{Graphite}}.
	\newblock \emph{\bibinfo{journal}{Phys. Rev. Lett.}}
	\textbf{\bibinfo{volume}{85}}, \bibinfo{pages}{5214--5217}
	(\bibinfo{year}{2000}).
	
	\bibitem{tan2002}
	\bibinfo{author}{Tan, P.} \emph{et~al.}
	\newblock \bibinfo{title}{Probing the phonon dispersion relations of graphite
		from the double-resonance process of stokes and anti-stokes raman scatterings
		in multiwalled carbon nanotubes}.
	\newblock \emph{\bibinfo{journal}{Phys. Rev. B}} \textbf{\bibinfo{volume}{66}},
	\bibinfo{pages}{245410} (\bibinfo{year}{2002}).
	
	\bibitem{cancado2002}
	\bibinfo{author}{Can\c{c}ado, L.~G.} \emph{et~al.}
	\newblock \bibinfo{title}{Stokes and anti-stokes double resonance raman
		scattering in two-dimensional graphite}.
	\newblock \emph{\bibinfo{journal}{Phys. Rev. B}} \textbf{\bibinfo{volume}{66}},
	\bibinfo{pages}{035415} (\bibinfo{year}{2002}).
	
	\bibitem{Marangoni2007}
	\bibinfo{author}{Marangoni, M.} \emph{et~al.}
	\newblock \bibinfo{title}{Narrow-bandwidth picosecond pulses by spectral
		compression of femtosecond pulses in a second-order nonlinear crystal}.
	\newblock \emph{\bibinfo{journal}{Opt. Express}} \textbf{\bibinfo{volume}{15}},
	\bibinfo{pages}{8884--8891} (\bibinfo{year}{2007}).
	
	\bibitem{PhysRevB.82.081408}
	\bibinfo{author}{Liu, W.-T.} \emph{et~al.}
	\newblock \bibinfo{title}{Nonlinear broadband photoluminescence of graphene
		induced by femtosecond laser irradiation}.
	\newblock \emph{\bibinfo{journal}{Phys. Rev. B}} \textbf{\bibinfo{volume}{82}},
	\bibinfo{pages}{081408} (\bibinfo{year}{2010}).
	
	\bibitem{Nair2008}
	\bibinfo{author}{Nair, R.~R.} \emph{et~al.}
	\newblock \bibinfo{title}{Fine {Structure} {Constant} {Defines} {Visual}
		{Transparency} of {Graphene}}.
	\newblock \emph{\bibinfo{journal}{Science}} \textbf{\bibinfo{volume}{320}},
	\bibinfo{pages}{1308--1308} (\bibinfo{year}{2008}).
	
	\bibitem{Pop2012}
	\bibinfo{author}{Pop, E.}, \bibinfo{author}{Varshney, V.} \&
	\bibinfo{author}{Roy, A.~K.}
	\newblock \bibinfo{title}{Thermal properties of graphene: Fundamentals and
		applications}.
	\newblock \emph{\bibinfo{journal}{MRS Bull.}} \textbf{\bibinfo{volume}{37}},
	\bibinfo{pages}{1273--1281} (\bibinfo{year}{2012}).
	
	\bibitem{Gironcoli1995}
	\bibinfo{author}{de~Gironcoli, S.}
	\newblock \bibinfo{title}{Lattice dynamics of metals from density-functional
		perturbation theory}.
	\newblock \emph{\bibinfo{journal}{Phys. Rev. B}} \textbf{\bibinfo{volume}{51}},
	\bibinfo{pages}{6773--6776} (\bibinfo{year}{1995}).
	
	\bibitem{Giannozzi2009}
	\bibinfo{author}{Giannozzi, P.} \emph{et~al.}
	\newblock \bibinfo{title}{Quantum espresso: a modular and open-source software
		project for quantum simulations of materials}.
	\newblock \emph{\bibinfo{journal}{J. Phys. Condens. Matter}}
	\textbf{\bibinfo{volume}{21}}, \bibinfo{pages}{395502}
	(\bibinfo{year}{2009}).
	
	\bibitem{wang_crystal_2012}
	\bibinfo{author}{Wang, Y.}, \bibinfo{author}{Panzik, J.~E.},
	\bibinfo{author}{Kiefer, B.} \& \bibinfo{author}{Lee, K. K.~M.}
	\newblock \bibinfo{title}{Crystal structure of graphite under room-temperature
		compression and decompression}.
	\newblock \emph{\bibinfo{journal}{Sci. Rep.}} \textbf{\bibinfo{volume}{2}}
	(\bibinfo{year}{2012}).
	
	\bibitem{Nef2014}
	\bibinfo{author}{Nef, C.} \emph{et~al.}
	\newblock \bibinfo{title}{High-yield fabrication of nm-size gaps in monolayer
		cvd graphene}.
	\newblock \emph{\bibinfo{journal}{Nanoscale}} \textbf{\bibinfo{volume}{6}},
	\bibinfo{pages}{7249--7254} (\bibinfo{year}{2014}).
	
\end{thebibliography}

\end{document}